\documentclass[a4paper,11pt]{article}
\pdfoutput=1 

\usepackage{jheppub} 

\usepackage[T1]{fontenc} 

\usepackage{etoolbox}

\title{\boldmath Yang-Baxter deformations of the $AdS_4\times\mathbb{CP}^3$ superstring sigma model}

\author{Ren\'e Negr\'on}
\author{and Victor O. Rivelles}
\affiliation{Instituto de F\'isica, Universidade de S\~ao Paulo,\\Rua do Mat\~ao, 1371, 05508-090, S\~ao Paulo, SP, Brazil}

\emailAdd{renenegronh@usp.br}
\emailAdd{rivelles@fma.if.usp.br}

\abstract{
The gravity dual of $\beta$-deformed ABJM theory can be obtained by a TsT transformation of $AdS_4\times\mathbb{CP}^3$. We present a supercoset construction of $\mathbb{CP}^3$ to obtain this gravity dual theory as a Yang-Baxter deformation. This is done by selecting a convenient combination of Cartan generators in order to get an Abelian $r$-matrix satisfying the classical Yang-Baxter equation. Our results provide another illustration of the  relation between Abelian $r$-matrices and TsT transformations.}

\begin{document} 

\makeatletter 
\patchcmd{\maketitle}{\@fpheader}{}{\hfill}{} 
\makeatother 

\maketitle
\flushbottom

\section{Introduction}
\label{intro}
	
Gauge/gravity duality is certainly one of the most important properties of systems involving quantum gravity. 
The original proposal establishes a correspondence between Type IIB superstring theory in $AdS_5\times S^5$ and $\mathcal{N}=4$ super-Yang-Mills theory in four dimensions \cite{Maldacena:1997re}. Also, there is no doubt that integrability has become one of the most useful tools to study both sides of the correspondence. It started when a Lax pair for the string theory side was constructed \cite{BPR1}. The $\mathbb{Z}_4$ automorphism of the $\mathfrak{psu}(2,2|4)$ algebra proved to be a key ingredient for integrability. However, a full proof of classical integrability, that is, the existence of an infinite number of conserved charges in involution, was only found later on \cite{Vicedo1,Magro1}. 

Integrability is not a very common property of two dimensional field theories. Furthermore, there is no systematic way to prove whether a theory is integrable or not. For this reason, relating the original superstring sigma model to deformed versions of it which still preserve the integrable structure is quite remarkable. A very important class of deformations, the so called $\beta$-deformations associated to a marginal deformation of $\mathcal{N}=4$ super-Yang-Mills theory, gives rise to strings moving in Lunin-Maldacena backgrounds \cite{LuninMaldacena}. These deformed models are also integrable \cite{Frolov2,Frolov3} and a systematic way of constructing the gravity dual of these gauge theories using TsT transformations was developed \cite{Imeroni:2008cr}.

Another important type of integrable deformation associated to string theory sigma models \cite{Vicedo6} makes use of $r$-matrices,  that is, linear operators satisfying the modified classical Yang-Baxter equation (mCYBE) \cite{Klimcik,Klimcik2}. The first kind of $r$-matrices that were used are of the Drinfield-Jimbo type and they allowed the determination of the background fields for these models \cite{Arutyunov:2013ega,Hoare:2014pna,Arutyunov:2015qva}. Integrable deformations can also be found if the $r$-matrices satisfy  the classical Yang-Baxter equation (CYBE) \cite{Kawaguchi:2014qwa}. In this case it is possible to find partial deformations, that is, either $AdS_5$ or $S^5$ is deformed in the $AdS_5\times S^5$ case. Furthermore, using these kind of deformations, it is possible to show that Lunin-Maldacena backgrounds have an algebraic origin \cite{Matsumoto:2014nra}\footnote{More recently, the relation between solutions of the CYBE and supergravity have been studied using open-closed string maps where Yang-Baxter deformations can be generalized beyond cosets \cite{Araujo:2017jkb,Bakhmatov:2017joy,Bakhmatov:2018apn}.}. In fact, the connection between Abelian $r$-matrices and TsT transformations for string theory sigma models has been observed in many situations \cite{Matsumoto:2014gwa,Matsumoto,Matsumoto:2014ubv,Matsumoto:2015uja,vanTongeren:2015soa,Matsumoto:2015ypa} and proven recently \cite{Osten:2016dvf}. Moreover, it has been suggested \cite{Hoare:2016wsk} and then demonstrated in general \cite{Borsato:2016pas,Borsato:2017qsx} that Yang-Baxter deformations are equivalent to non-Abelian duality transformations for superstring sigma models.


Another well known gauge/gravity duality is for the $AdS_4/CFT_3$ case which establishes a correspondence between Type IIA superstring theory in $AdS_4\times \mathbb{CP}^3$ and $\mathcal{N}=6$ superconformal Chern-Simons-matter theory \cite{Aharony:2008ug}. Many properties of superstrings in $AdS_5\times S^5$ can be extended to the $AdS_4\times\mathbb{CP}^3$ counterpart. In particular, both of them admit a supercoset description with $\mathbb{Z}_4$ grading so that many integrability aspects are similar \cite{Arutyunov:2008if}. As a consequence, all of the analysis that lead to the formulation of Yang-Baxter deformations in $AdS_5\times S^5$ can in principle be extended to $AdS_4\times\mathbb{CP}^3$. Evidence for this is the fact that TsT transformations can also be used to construct the gravity dual of $\beta$-deformations of the ABJM theory \cite{Imeroni:2008cr}. The present work is devoted to show that  Yang-Baxter deformations can also be used to construct these deformed backgrounds thus exhibiting their algebraic $r$-matrix origin.

The usual coset description of $AdS_4\times \mathbb{CP}^3$ is given by
\begin{equation}
\frac{OSp(2,2|6)}{SO(1,3)\times U(3)}\,.
\label{AdS4CP3coset}
\end{equation}
However, the $\mathbb{CP}^3$ metric that is obtained from this coset does not exactly match the standard Fubini-Study metric \cite{Klose:2010ki}. Furthermore, the parametrization using  generators of $SO(6)$ which are $6\times 6$ matrices turns out to be very intricate turning computations quite messy. The solution  proposed in this paper is to modify the coset describing the $\mathbb{CP}^3$ part of the sigma model to 
\begin{equation}
\frac{SU(4)\times SU(2)}{U(3)\times SU(2)}\,.
\end{equation}
This description is analogous to the one performed in the analysis of Yang-Baxter deformations of $T^{1,1}$ \cite{Crichigno:2014ipa}. As we will show, it is possible to embed some of the generators of $\mathfrak{u}(3)\oplus\mathfrak{su}(2)$ into $\mathfrak{su}(4)\oplus\mathfrak{su}(2)$, giving rise to an orthogonal basis that can be used to get the standard Fubini-Study metric of $\mathbb{CP}^3$.

This paper is organized as follows. In section 2 we show how to obtain the standard $\mathbb{CP}^3$ metric as discussed above.  
In section 3 we present the appropriate combinations of Cartan generators needed to get the Yang-Baxter deformation of our coset.  Finally, in section 4, we present some concluding remarks. Appendix A reviews the usual coset construction of $\mathbb{CP}^3$ while in Appendix B we present our conventions and collect useful formulae for $\mathfrak{su}(4)$.

\section{Coset construction of $\mathbb{CP}^3$} 
\label{section2}

There are only two superstring sigma models with vanishing $\beta$-function in ten dimensional semi-symmetric spaces, $AdS_5\times S^5$ and $AdS_4\times \mathbb{CP}^3$ \cite{Zarembo:2010sg}. Like $AdS_5\times S^5$, $AdS_4\times \mathbb{CP}^3$ is classically integrable and admits a coset formulation \cite{Arutyunov:2008if,Stefanski:2008ik,Sorokin:2010wn}. In this section we will briefly review some of its main properties.

\subsection{Superstrings in $AdS_4\times\mathbb{CP}^3$}

Type IIA superstrings moving in $AdS_4\times\mathbb{CP}^3$ can be described by a sigma model formulated in the coset (\ref{AdS4CP3coset}). Its Lagrangian is a straightforward extension of the Metsaev-Tseytlin Lagrangian and it is given by \cite{Arutyunov:2008if}
\begin{equation}
\mathcal{L} = -\frac{R^2}{4\pi\alpha'}\,\mbox{STr}\left[\gamma^{\alpha\beta}A^{(2)}_\alpha A^{(2)}_\beta-\epsilon^{\alpha\beta}A^{(1)}_\alpha A^{(3)}_\beta\right]\,, \label{2.1}
\end{equation}
where $\alpha$, $\beta$ are world-sheet indices, the metric and the Levi-Civita tensor are $\gamma^{\alpha\beta}=\mbox{diag}(-1,1)$ and $\epsilon^{\alpha\beta}=1$, respectively, and $R$ is the radius of  $\mathbb{CP}^3$. $A_\alpha=g^{-1}\partial_\alpha g$, with $g\in OSp(2,2|6)$, is the usual Maurer-Cartan current. 

One important and subtle aspect of this formulation is that the supercoset (\ref{AdS4CP3coset}) only has 24 fermionic directions which is the number of supersymmetries preserved by the background. In ten dimensions, the Green-Schwarz superstring is built using two Majorana-Weyl fermions with a total of 32 degrees of freedom. Thus, the coset description of $AdS_4\times\mathbb{CP}^3$ is missing 8 fermions and,  therefore, lacks the correct amount of supersymmetry of usual GS superstrings. We can go  around this problem arguing that the 8 missing fermions are part of the 16 fermionic degrees of freedom that  are removed by $\kappa$-symmetry. In this way the coset (\ref{AdS4CP3coset}) can be thought of as a model with partially fixed $\kappa$-symmetry \cite{Arutyunov:2008if,Gomis:2008jt}.

Let us consider the geometrical structure of $AdS_4\times\mathbb{CP}^3$. The metric can be split in two terms
\begin{equation}
ds^2 = R^2\left[\frac{1}{4}\,ds^2_{AdS_4}+ds^2_{\mathbb{CP}^3}\right]\,,
\end{equation}
with the factor $1/4$ being due to the fact that the radius of $\mathbb{CP}^3$ is twice the radius of $AdS_4$. In global coordinates the metric for $AdS_4$ reads
\begin{eqnarray}
ds^2_{AdS_4} = -\cosh^2 \rho\, dt^2+d\rho^2+\sinh^2\rho\left(d\theta^2+\sin^2\theta\, d\varphi^2\right)\,,
\end{eqnarray}
where  $-\infty<t<\infty$, $0\le \rho<\infty$, $0\le\theta<\pi$ and $0\le\varphi<2\pi$. The standard Fubini-Study metric on $\mathbb{CP}^3$ is
\begin{eqnarray}
ds^2_{\mathbb{CP}^3} = && d\xi^2+\frac{1}{4}\cos^2\xi\left(d\theta^2_1+\sin^2\theta_1 d\varphi^2_1\,\right)+\frac{1}{4}\sin^2\xi\left(d\theta^2_2+\sin^2\theta_2 d\varphi^2_2\,\right) \nonumber \\
&& +\cos^2\xi\sin^2\xi\left(\frac{1}{2}\,\cos\theta_1 d\,\varphi_1-\frac{1}{2}\,\cos\theta_2 d\,\varphi_2+d\psi\right)^2\,,
\label{CPMetricNonDeformed}
\end{eqnarray}
where $(\theta_1,\varphi_1)$ and $(\theta_2,\varphi_2)$ parametrize two spheres, the angle $0\le\xi<\pi/2$ determines their radii, and the angle $0\le\psi<2\pi$ corresponds to a $U(1)$ isometry.

\subsection{Supercoset construction of $\mathbb{CP}^3$} 

As we have discussed in the introduction, the usual coset representation of $\mathbb{CP}^3$ is not convenient to formulate a Yang-Baxter sigma model. Instead let us consider the following coset\footnote{A similar construction was done for Yang-Baxter deformations of $T^{1,1}$ \cite{Crichigno:2014ipa}.}
\begin{equation}
\mathbb{CP}^3 = \frac{SU(4)\times SU(2)}{U(3)\times SU(2)}\,,
\label{AlterCoset}
\end{equation}
The basis of $\mathfrak{su}(4)\oplus\mathfrak{su}(2)$ that we will consider is composed of $\mathfrak{su}(4)$ generators denoted by $\mathbf{L}_m\,\,\,(m=1,...\,,15)$ and $\mathfrak{su}(2)$ generators denoted by $\mathbf{M}_{a}\,\,\,(a=1,2,3)$. As we have discussed, the usual construction of $\mathbb{CP}^3$ as a coset involves a representation of the generators as $6\times 6$ bosonic matrices. Instead of that we will choose $(4|2)\times(4|2)$ supermatrices in the fundamental representation of $\mathfrak{su}(4)\oplus\mathfrak{su}(2)$ 
\begin{equation}
\mathbf{L}_m =-\frac{i}{2}\left(\begin{array}{c|c}
\lambda_m & \mathbf{0}_{4\times 2} \\
\hline
\mathbf{0}_{2\times 4} & \mathbf{0}_{2\times 2}
\end{array}\right)\,,\qquad
\mathbf{M}_a =-\frac{i}{2}\left(\begin{array}{c|c}
\mathbf{0}_{4\times 4} & \mathbf{0}_{4\times 2} \\
\hline
\mathbf{0}_{2\times 4} & \sigma_a
\end{array}\right)\,,
\end{equation}
where $\lambda_m$ are the conventional $4\times 4$ $\mathfrak{su}(4)$ generators generalizing the Gell-Mann matrices of $\mathfrak{su}(3)$\footnote{For a complete review of the representations of $\mathfrak{su}(4)$ see \cite{Pfeifer}.}, while $\sigma_a$ are the usual $2\times 2$ Pauli matrices.
The commutation relations and the supertraces 
are 
\begin{eqnarray}
\left[\mathbf{L}_m,\mathbf{L}_n\right] = f^{\,\,\,\,\,\,\,p}_{mn}\,\mathbf{L}_{p}\,&,& \quad \left[\mathbf{M}_a,\mathbf{M}_b\right] = \epsilon^{\,\,\,\,\,\,\,c}_{ab}\,\mathbf{M}_c\,, \\
\mbox{STr}(\mathbf{L}_m\,\mathbf{L}_n) = -\frac{1}{2}\delta_{mn}\,&,& \quad \mbox{STr}(\mathbf{M}_a\,\mathbf{M}_b) = \frac{1}{2}\delta_{ab}\,, \label{STRLM}
\end{eqnarray}
where $f^{\,\,\,\,\,\,\,p}_{mn}$ are the structure constants of $\mathfrak{su}(4)$ while for $\mathfrak{su}(2)$ we normalize $\epsilon_{123}=1$. We notice that the Cartan generators of $\mathfrak{su}(4)\oplus\mathfrak{su}(2)$ are given by $\mathbf{L}_3, \mathbf{L}_8, \mathbf{L}_{15}$ and $\mathbf{M}_3$. We will make use of these generators to construct $r$-matrices in the next section.


We will now show that our coset construction (\ref{AlterCoset}) does indeed reproduce the $\mathbb{CP}^3$ metric. In order to get the Fubini-Study metric for  $\mathbb{CP}^3$ we first redefine some of the generators of the basis of $\mathfrak{su}(4)\oplus\mathfrak{su}(2)$. The following combination of the generators $\mathbf{L}_6\subset \mathfrak{u}(3)$ and $\mathbf{M}_1\subset \mathfrak{su}(2)$ 
turns out to be very convenient
\begin{equation}
\mathbf{T}_1=\mathbf{L}_6-\mathbf{L}_9\,,\qquad \mathbf{T}_2 = \mathbf{L}_6+\mathbf{L}_9+2\mathbf{M}_1\,.
\end{equation}
The coefficients of $\mathbf{T}_2$ have been fixed to reproduce exactly the $\mathbb{CP}^3$ metric. 
The orthogonal basis for $\mathfrak{su}(4)\oplus\mathfrak{su}(2)$ after this embedding is 
\begin{equation}
\mathfrak{su}(4)\oplus\mathfrak{su}(2)= \mbox{span}_{\mathbb{R}}\{\mathbf{L}_{m'},\mathbf{M}_2,\mathbf{M}_3,\mathbf{T}_1,\mathbf{T}_2,\mathbf{H}\}\,,
\label{basisT1T2}
\end{equation}
where $\mathbf{L}_{m'}$ runs through all the basis of $\mathfrak{su}(4)$ except for $\mathbf{L}_6$ and $\mathbf{L}_9$, and  $\mathbf{H}$ is defined as
\begin{equation}
\mathbf{H} = \mathbf{L}_6+\mathbf{L}_9+\mathbf{M}_1\,,
\end{equation}
with norm 
\begin{equation}
\mbox{STr}(\mathbf{H}\,\mathbf{H}) = -\frac{1}{2}\,.
\label{STRH}
\end{equation}
With these redefinitions, and in order to write an appropriate coset parametrization of (\ref{AlterCoset}), we consider the following orthogonal basis
\begin{equation}
\frac{\mathfrak{su}(4)\oplus\mathfrak{su}(2)}{\mathfrak{u}(3)\oplus\mathfrak{su}(2)} = \mbox{span}_{\mathbb{R}}\{\mathbf{P}_m\}\,, \qquad m=1,\,...\,,6\,,
\label{basis}
\end{equation}
where, for the sake of simplicity, we have renamed the generators as
\begin{eqnarray}
\mathbf{P}_1 = \mathbf{L}_{11}\,, \qquad \mathbf{P}_2 =& \mathbf{L}_{12}\,, \qquad \mathbf{P}_3 &= \mathbf{L}_{13}\,, \nonumber \\
\mathbf{P}_4 = \mathbf{L}_{14}\,, \qquad \mathbf{P}_5 =& \mathbf{H}\,, \qquad \mathbf{P}_6 &= \mathbf{L}_{10}\,.
\label{DefPm}
\end{eqnarray}
Using (\ref{STRLM}) and (\ref{STRH}) we find
\begin{equation}
\mbox{STr}(\mathbf{P}_m\,\mathbf{P}_n) = -\frac{1}{2}\delta_{mn}\,.
\label{STrPm}
\end{equation}
As it can be seen from (\ref{basis}), the $\mathfrak{u}(3)$ generators are $\mathbf{L}_1,\mathbf{L}_2,\mathbf{L}_3,\mathbf{L}_4,\mathbf{L}_5,\mathbf{L}_7,\mathbf{L}_8,\mathbf{T}_1,\mathbf{L}_{15}$ while the $\mathfrak{su}(2)$ generators are $\mathbf{T}_2,\mathbf{M}_2,\mathbf{M}_3$.  We will use the basis (\ref{basis}) to parametrize $\mathbb{CP}^3$.

We can now write an appropriate coset representative which will allow us to get the desired $\mathbb{CP}^3$ metric. It is given by
\begin{eqnarray}
g = \exp{\left[\varphi_1\mathbf{L}_3+\varphi_2 \mathbf{L}-\psi\mathbf{M}_3\right]}\exp{\left[\theta_1 \mathbf{L}_2+(\theta_2+\pi)\mathbf{L}_{14}\right]}\exp{\left[(2\rho+\pi)(\mathbf{L}_{10}+\mathbf{M}_2)\right]}\,,
\label{gCP3}
\end{eqnarray}
where $\mathbf{L}$ is given by  
\begin{equation}
\mathbf{L}=-\frac{1}{\sqrt{3}}\mathbf{L}_8+\sqrt{\frac{{2}}{{3}}}\mathbf{L}_{15}\,.
\label{LDefinition}
\end{equation}
Then we can write a Lie algebra element of $\mathfrak{su}(4)\oplus\mathfrak{su}(2)$ by using the Maurer-Cartan one form $A=g^{-1}d\,g$.

In order to write the metric we need to project out all of the components proportional to generators of $\mathfrak{u}(3)\oplus\mathfrak{su}(2)$ in the Maurer-Cartan form. To this end let us define the projector
\begin{eqnarray}
P(A) &=& A+2\,\Big[\sum_{i}\mathbf{L}_{i}\,\mbox{STr}(\mathbf{L}_i A)-\frac{\mathbf{T}_1}{2}\,\mbox{STr}(\mathbf{T}_1A)+\frac{\mathbf{T}_2}{2}\,\mbox{STr}(\mathbf{T}_2A) \nonumber \\
&& \quad\qquad-\mathbf{M}_2\,\mbox{STr}(\mathbf{M}_2A)-\mathbf{M}_3\,\mbox{STr}(\mathbf{M}_3A)\Big] \nonumber \\
&=& \cos\xi\sin\theta_1 d\varphi_1\,\mathbf{P}_1+\cos\xi d\theta_1\,\mathbf{P}_2-\sin\theta_2\sin\xi d\varphi_2\,\mathbf{P}_3 \label{Projector1}  \\
&& -\sin\xi d\theta_2\,\mathbf{P}_4-\sin{\xi}\cos\xi\left(\cos\theta_1 d\varphi_1-\cos\theta_2 d\varphi_2+2 d\psi\right)\mathbf{P}_5+2d\xi\,\mathbf{P}_6\,, \nonumber
\end{eqnarray}
where in the first line the summation runs over $i=1,2,3,4,5,7,8,15$. Then the metric for $\mathbb{CP}^3$ (\ref{CPMetricNonDeformed}) can be obtained from the following symmetric Lagrangian
\begin{equation}
\mathcal{L} = -\frac{1}{2}\,\gamma^{\alpha\beta}\,\mbox{STr}\left[A_\alpha\,P(A_\beta)\right]\,.
\label{CP3Lagrangian}
\end{equation}

\section{Yang-Baxter deformations of $AdS_4\times\mathbb{CP}^3$}

One of the most interesting techniques to deform sigma models is the $\eta$-deformation based on $r$-matrices satisfying the Yang-Baxter equation. It was originally developed for the principal chiral model \cite{Klimcik,Klimcik2} and then generalised to coset sigma models \cite{Vicedo5}. It should be noticed that this deformation can also be applied to non integrable field theories \cite{Crichigno:2014ipa}. This deformation technique can  be straightforwardly applied to type IIA superstrings moving in $AdS_4\times\mathbb{CP}^3$ because it has the same structure as the Metsaev-Tseytlin superstring in $AdS_5\times S^5$. This will be done in this section.
%

\subsection{Yang-Baxter sigma model for $\mathbb{CP}^3$}

The $\beta$ deformed background of \cite{Imeroni:2008cr} is a three parameter  deformation of the $\mathbb{CP}^3$ part of $AdS_4\times\mathbb{CP}^3$.
In the context of Yang-Baxter deformations this is  possible only for $r$-matrices satisfying the CYBE \cite{Kawaguchi:2014qwa,Matsumoto:2015jja}. So we will deform the bosonic part of the Lagrangian (\ref{2.1}) based on the classical Yang-Baxter equation in order to get the metric and the $B$ field. The deformed Lagrangian is \cite{Matsumoto:2015jja}
\begin{equation}
\mathcal{L}_\eta = -\frac{1}{2}(\gamma^{\alpha\beta}-\epsilon^{\alpha\beta})\,\mbox{STr}\left(A_{\alpha} P\circ\mathcal{O}^{-1}\,A_\beta\right)\,,
\label{DefLagran}
\end{equation}
where the operator $\mathcal{O}$, that depends on the deformation parameter $\eta$, is given by
\begin{equation}
\mathcal{O}^{-1}=\frac{1}{1-2\eta R_g\circ P}\,.
\label{DefOperator}
\end{equation}
The projector $P$ is given in (\ref{Projector1}) while $R_g$ is 
\begin{equation}
R_g(X) = \mbox{Ad}_{g^{-1}}\circ R\circ \mbox{Ad}_g\,(X)\,,
\label{RgDefinitionCP}
\end{equation}
where $g$ is a coset representative, and the linear operator $R$ is an $r$-matrix satisfying the classical Yang-Baxter equation. This operator can also be written in tensorial notation as 
\begin{equation}
R(X) = \mbox{STr}_2[r(1\otimes X)] = \sum_i\left[a_i\mbox{STr}(b_iX)-b_i\mbox{STr}(a_iX)\right]\,,
\label{RXDefi}
\end{equation}
where $a_i$ and $b_i$ denote elements of the basis for $\mathfrak{su}(4)\oplus\mathfrak{su}(2)$ and the $r$-matrix is constructed using these generators as
\begin{equation}
r = \sum_i a_i\wedge b_i = \sum_i\left(a_i\otimes b_i-b_i\otimes a_i\right)\,.
\end{equation}
In order to extract the background fields from the Lagrangian (\ref{DefLagran}) we need to find the action of $R_g$ on each generator of the basis (\ref{basis}). This can be written as
\begin{equation}
R_g(\mathbf{P}_m) = \Lambda^{\,\,\,n}_{m}\,\mathbf{P}_n+\,...\,,
\label{Racting}
\end{equation}
where $\mathbf{P}_m$ are defined in (\ref{DefPm}) and the dots stand for generators of $\mathfrak{u}(3)\oplus\mathfrak{su}(2)$ which can be ignored because they will be projected out by $P$.

The action of the operator (\ref{DefOperator}) can also be computed in a similar way 
\begin{equation}
\mathcal{O}^{-1}(\mathbf{P}_m) = K^{\,\,\,n}_{m}\,\mathbf{P}_n+\,...\,,
\label{Oacting}
\end{equation}
where again the dots represent terms that will be projected out. Combining equations (\ref{Racting}) and (\ref{Oacting}) it is easy to find the relation between the coefficients $\Lambda^{\,\,\,n}_m$ and $K^{\,\,\,n}_m$
\begin{equation}
\mathbf{P}_m = \left(K^{\,\,\,n}_{m}\,\mathbf{P}_{n}-2\,\eta\,K^{\,\,\,n}_{m}\Lambda^{\,\,\,p}_n\,\mathbf{P}_p\right)\,,
\end{equation}
which can be solved for $K^{\,\,\,n}_{m}$. In matrix notation the solution is 
\begin{equation}
\mathbf{K} = \left(\mathbf{I}-2\,\eta\,\mathbf{\Lambda}\,\right)^{-1}\,.
\label{KMatrix}
\end{equation}
The Maurer-Cartan current $A_\alpha$ can now be expanded in the basis (\ref{basisT1T2}). However, as we have discussed in section \ref{section2}, the operator $P$ will project the current to the basis (\ref{basis}) giving\footnote{We should stress that despite of the notation being used $E^m_\alpha$ does not represent the usual vielbein of $\mathbb{CP}^3$ but rather the coefficients in front of each of the generators in the third and fourth lines of (\ref{Projector1}).}
\begin{equation}
P(A_\alpha) = E^{m}_{\alpha}\,\mathbf{P}_m\,.
\label{PAction}
\end{equation}
Using  (\ref{Oacting}) and (\ref{PAction}) in (\ref{DefLagran}) we finally get
\begin{equation}
\mathcal{L}_{\eta} 
=\frac{1}{4}\left(\gamma^{\alpha\beta}K_{(mn)}E^m_\alpha E^n_\beta-\epsilon^{\alpha\beta}K_{[mn]}E^m_\alpha E^n_\beta\right),
\label{GandBfield}
\end{equation}
where the coefficients $K_{(mn)}$ and $K_{[mn]}$ are the symmetric and antisymmetric parts of the matrix (\ref{KMatrix}). The metric and the antisymmetric field can then  easily be read off from (\ref{GandBfield}). 

Finally, the last step is the choice of an $r$-matrix. The most natural choice are the so called Abelian $r$-matrices which are basically constructed out of Cartan generators. There are four Cartan generators in $\mathfrak{su}(4)\oplus\mathfrak{su}(2)$, that is, $\mathbf{L}_3, \mathbf{L}_8, \mathbf{L}_{15}$ and $\mathbf{M}_3$. In order to get a three parameter deformation we make the choice
\begin{equation}
r = \mu_1\,\mathbf{L}\wedge \mathbf{M}_3+\mu_2\,\mathbf{L}_3\wedge \mathbf{M}_3+\mu_3\,\mathbf{L}_3\wedge \mathbf{L}\,,
\label{RCP3_3P}
\end{equation}
where $\mathbf{L}$ is a linear combination of Cartan generators defined in (\ref{LDefinition}) and $\mu_1,\mu_2$ and $\mu_3$ are arbitrary parameters. The action of the operator $R_g$ on each of the generators (\ref{basis}) is given by (\ref{Racting}). Using the coset parametrization (\ref{gCP3}) together with definitions (\ref{RgDefinitionCP}) and (\ref{RXDefi}) it can be shown that the only non-vanishing components of the deformation matrix $\Lambda^{\,\,\,n}_m$ are 
\begin{eqnarray}
\Lambda^{\,\,\,3}_{1} = -\Lambda^{\,\,\,1}_{3} &=& -\frac{1}{2}\,\mu_3\sin\theta_1\sin\theta_2\sin\xi\cos\xi\,, \nonumber \\
\Lambda^{\,\,\,5}_{1} = -\Lambda^{\,\,\,1}_5   &=& -\frac{1}{2}\,(2\mu_2+\mu_3\cos\theta_2)\cos^2\xi\sin\theta_1\sin\xi\,, \nonumber \\
\Lambda^{\,\,\,5}_{3} = -\Lambda^{\,\,\,3}_5   &=&  \frac{1}{2}\,(2\mu_1+\mu_3\cos\theta_1)\cos\xi\sin\theta_2\sin^2\xi\,,
\end{eqnarray}
while the non-vanishing elements of $K^{\,\,\,n}_m$ are
\begin{eqnarray}
K^{\,\,\,1}_{1} &=& \mathcal{M}\,\big[1+\eta^2(2\mu_1+\mu_3\cos\theta_1)^2\cos^2\xi\sin^2\theta_2\sin^4\xi\big]\,, \nonumber \\
K^{\,\,\,1}_3 &=& \mathcal{M}\,\big[-\eta^2(2\mu_1+\mu_3\cos\theta_1)(2\mu_2+\mu_3\cos\theta_2)\cos^3\xi\sin\theta_1\sin\theta_2\sin^3\xi+ \nonumber \\
&& +\frac{1}{2}\mu_3\eta\sin\theta_1\sin\theta_2\sin 2\xi\big]\,, \nonumber \\
K^{\,\,\,1}_5 &=& \mathcal{M}\,\big[\eta\,(2\mu_2+\mu_3\cos\theta_2)\cos^2\xi\sin\theta_1\sin\xi+ \nonumber \\
&& -\frac{1}{2}\mu_3\eta^2(2\mu_1+\mu_3\cos\theta_1)\cos\xi\sin\theta_1\sin^2\theta_2\sin^2\xi\sin 2\xi\big]\,, \nonumber \\
K^{\,\,\,3}_1 &=& \mathcal{M}\,\big[\eta^2(2\mu_1+\mu_3\cos\theta_1)(2\mu_2+\mu_3\cos\theta_2)\cos^3\xi\sin\theta_1\sin\theta_2\sin^3\xi+ \nonumber \\
&& -\frac{1}{2}\mu_3\eta\sin\theta_1\sin\theta_2\sin 2\xi\big]\,, \nonumber \\
K^{\,\,\,3}_3 &=& \mathcal{M}\,\big[1+\eta^2(2\mu_2+\mu_3\cos\theta_2)^2\cos^4\xi\sin^2\theta_1\sin^2\xi\big]\,, \nonumber \\
K^{\,\,\,3}_5 &=& \mathcal{M}\,\big[-\eta(2\mu_1+\mu_3\cos\theta_1)\cos\xi\sin\theta_2\sin^2\xi \nonumber \\
&& -\frac{1}{2}\mu_3\eta^2(2\mu_2+\mu_3\cos\theta_2)\cos^2\xi\sin^2\theta_1\sin\theta_2\sin\xi\sin 2\xi\big]\,, \nonumber \\
K^{\,\,\,5}_1 &=& \mathcal{M}\,\big[-\eta\,(2\mu_2+\mu_3\cos\theta_2)\cos^2\xi\sin\theta_1\sin\xi+ \nonumber \\
&& -\frac{1}{2}\mu_3\eta^2(2\mu_1+\mu_3\cos\theta_1)\cos\xi\sin\theta_1\sin^2\theta_2\sin^2\xi\sin 2\xi\big]\,, \nonumber \\
K^{\,\,\,5}_3 &=& \mathcal{M}\,\big[\eta\,(2\mu_1+\mu_3\cos\theta_1)\cos\xi\sin\theta_2\sin^2\xi \nonumber \\
&& -\frac{1}{2}\mu_3\eta^2(2\mu_2+\mu_3\cos\theta_2)\cos^2\xi\sin^2\theta_1\sin\theta_2\sin\xi\sin 2\xi\big]\,, \nonumber \\
K^{\,\,\,5}_5 &=& \mathcal{M}\,\big[1+\frac{1}{4}\,\mu^2_3\,\eta^2\sin^2\theta_1\sin^2\theta_2\sin^2 2\xi\big]\,, \nonumber \\
K^{\,\,\,2}_2 &=& K^{\,\,\,4}_4 = K^{\,\,\,6}_6 = 1\,,
\end{eqnarray}
where
\begin{eqnarray}
\mathcal{M}^{-1} &=& 1+\eta^2\cos^2\xi\sin^2\xi\,\Big[\mu_3^2\sin^2\theta_1\sin^2\theta_2+(2\mu_2+\mu_3\cos\theta_2)^2\cos^2\xi\sin^2\theta_1+ \nonumber \\
&& +(2\mu_1+\mu_3\cos\theta_1)^2\sin^2\xi\sin^2\theta_2\Big]\,.
\end{eqnarray}
Using these results in (\ref{GandBfield}) we can find  the symmetric and antisymmetric parts of the deformed Lagrangian 
\begin{eqnarray}
L_{sym} &=& \gamma^{\alpha\beta}\mathcal{M}\bigg[\frac{1}{4}\cos^2\xi\left(\mathcal{M}^{-1}\partial_\alpha\theta_1\partial_\beta\theta_1+\sin^2\theta_1 \partial_\alpha\varphi_1\partial_\beta\varphi_1\,\right) \nonumber \\
&& \quad\quad+\frac{1}{4}\sin^2\xi\left(\mathcal{M}^{-1}\partial_\alpha\theta_2\partial_\beta\theta_2+\sin^2\theta_2 \partial_\alpha\varphi_2\partial_\beta\varphi_2\,\right) \nonumber \\
&& \quad\quad+\cos^2\xi\sin^2\xi\left(\partial_\alpha\psi+\frac{1}{2}\cos\theta_1 \partial_\alpha\varphi_1-\frac{1}{2}\cos\theta_2 \partial_\alpha\varphi_2\right) \nonumber \\
&& \quad\quad\times\left(\partial_\beta\psi+\frac{1}{2}\cos\theta_1 \partial_\beta\varphi_1-\frac{1}{2}\cos\theta_2 \partial_\beta\varphi_2\right)+\mathcal{M}^{-1}\partial_\alpha\xi\partial_\beta\xi \nonumber \\
&& +\eta^2\,\sin^4\xi\cos^4\xi\sin^2\theta_1\sin^2\theta_2\left(-\mu_1\partial_\alpha\varphi_1+\mu_2\partial_\alpha\varphi_2+\mu_3\partial_\alpha\psi\right) \nonumber \\
&& \quad\quad\times\left(-\mu_1\partial_\beta\varphi_1+\mu_2\partial_\beta\varphi_2+\mu_3\partial_\beta\psi\right)\bigg]\,, 
\end{eqnarray}
\begin{eqnarray}
L_{anti} &=& -2\,\epsilon^{\alpha\beta}\eta\,\mathcal{M}\sin^2\xi\cos^2\xi\,\bigg[\frac{1}{2}(2\mu_2+\mu_3\cos\theta_2)\cos^2\xi\sin^2\theta_1\,\partial_\alpha\psi\partial_\beta\varphi_1 \nonumber \\
&& +\frac{1}{2}(2\mu_1+\mu_3\cos\theta_1)\sin^2\xi\sin^2\theta_2\,\partial_\alpha\psi\partial_\beta\varphi_2 \nonumber \\
&& +\frac{1}{4}\Big(\mu_3\sin^2\theta_1\sin^2\theta_2+(2\mu_2+\mu_3\cos\theta_2)\cos^2\xi\sin^2\theta_1\cos\theta_2 \nonumber \\
&& (2\mu_1+\mu_3\cos\theta_1)\sin^2\xi\sin^2\theta_2\cos\theta_1\Big)\,\partial_\alpha\varphi_1\partial_\beta\varphi_2\bigg]\,.
\end{eqnarray}
If we define 
\begin{equation}
\hat{\gamma}_1= -\mu_1\eta \,, \qquad \hat{\gamma}_2=\mu_2\eta  \,, \qquad \hat{\gamma}_3= \mu_3\eta \,,
\end{equation}
we can read the metric and antisymmetric fields
\begin{eqnarray}
ds^2_{\mathbb{CP}^3_{\hat{\gamma}}} &=& d\xi^2+\frac{1}{4}\cos^2\xi\left(d\theta^2_1+\mathcal{M}\sin^2\theta_1 d\varphi^2_1\,\right)+\frac{1}{4}\sin^2\xi\left(d\theta^2_2+\mathcal{M}\sin^2\theta_2 d\varphi^2_2\,\right)+ \nonumber \\
&& +\,\mathcal{M}\cos^2\xi\sin^2\xi\left(d\psi+\frac{1}{2}\cos\theta_1 d\varphi_1-\frac{1}{2}\cos\theta_2 d\varphi_2\right)^2+ \nonumber \\
&& +\mathcal{M}\sin^4\xi\cos^4\xi\sin^2\theta_1\sin^2\theta_2\left(\hat{\gamma}_1 d\varphi_1+\hat{\gamma}_2 d\varphi_2+\hat{\gamma}_3 d\psi\right)^2\,,\\
&& \nonumber \\
B &=& -\mathcal{M}\sin^2\xi\cos^2\xi\,\bigg[\frac{1}{2}(2\hat{\gamma}_2+\hat{\gamma}_3\cos\theta_2)\cos^2\xi\sin^2\theta_1\,d\psi\wedge d\varphi_1+ \nonumber \\
&& +\frac{1}{2}(-2\hat{\gamma}_1+\hat{\gamma}_3\cos\theta_1)\sin^2\xi\sin^2\theta_2\,d\psi\wedge d\varphi_2+ \nonumber \\
&& +\frac{1}{4}\Big(\hat{\gamma}_3\sin^2\theta_1\sin^2\theta_2+(2\hat{\gamma}_2+\hat{\gamma}_3\cos\theta_2)\cos^2\xi\sin^2\theta_1\cos\theta_2+ \nonumber \\
&& (-2\hat{\gamma}_1+\hat{\gamma}_3\cos\theta_1)\sin^2\xi\sin^2\theta_2\cos\theta_1\Big)\,d\varphi_1\wedge d\varphi_2\bigg]\,,
\end{eqnarray}
which are in complete agreement with the three-parameter deformation of $AdS_4\times\mathbb{CP}^3$ which was obtained using TsT transformations \cite{Imeroni:2008cr}. As a consequence we have shown that the three-parameter deformation arises naturally from Yang-Baxter deformations having an algebraic origin given by the $r$-matrix (\ref{RCP3_3P}). 

\section{Concluding remarks}

In this paper we have derived the metric and $B$ field of $\beta$ deformed type IIA $AdS_4\times\mathbb{CP}^3$ superstring theory as a Yang-Baxter deformation. 
In order to do that, a new supercoset description of $\mathbb{CP}^3$ was proposed. The main advantage of this parametrization is that it leads to the standard Fubini-Study metric of $\mathbb{CP}^3$ and makes Yang-Baxter deformations more tractable.
It can be regarded as a new non-trivial example of the relation between TsT transformations and solutions of the classical Yang-Baxter equation. 

The procedure developed here can also be applied to other situations. As it is well known, Yang-Baxter deformations lead to supergravity backgrounds only if the associated $r$-matrix is unimodular \cite{Borsato:2016ose}. Abelian $r$-matrices trivially satisfy the unimodularity condition and they are related to TsT transformations \cite{Osten:2016dvf}. In particular, the gravity dual of canonical non-commutative super Yang-Mills theory can also be obtained using Abelian $r$-matrices \cite{Matsumoto:2014gwa}. On the other hand, non-Abelian $r$-matrices can be unimodular or not. Yang-Baxter deformations related to unimodular non-Abelian $r$-matrices have been interpreted as the gravity dual of some non-commutative gauge theory \cite{vanTongeren:2015uha,vanTongeren:2016eeb}. The gravitational duals of non-commutative and dipole deformations of ABJM theory found in \cite{Imeroni:2008cr} can, in principle, be obtained as Yang-Baxter deformations. In particular, an extension of our coset parametrization could be built to include deformations of the $AdS_4$ part in order to compute dipole deformations. Apart from that, as it was shown in the supercoset construction of $AdS_5\times S^5$ \cite{Arutyunov:2015qva,Kyono:2016jqy}, it seems possible to  generalize our results to reproduce the two and four form  Ramond-Ramond fluxes that appear in $\beta$ deformed $AdS_4\times \mathbb{CP}^3$ but we will leave that for a future work.


\appendix

\section{Usual parametrization of $\mathbb{CP}^3$}


The usual parametrization of $\mathbb{CP}^3$ can be given in terms of the following six-parametric matrix \cite{Arutyunov:2008if,Stefanski:2008ik}
\begin{equation}
Y = \left(\begin{array}{rrrrrr}
0 & 0 & y_1 & y_2 & y_3 & y_4 \\
0 & 0 & y_2 & -y_1 & y_4 & -y_3 \\
-y_1 & -y_2 & 0 & 0 & y_5 & y_6 \\
-y_2 & y_1 & 0 & 0 & y_6 & -y_5 \\
-y_3 & -y_4 & -y_5 & -y_6 & 0 & 0 \\
-y_4 & y_3 & -y_6 & y_5 & 0 & 0 \\
\end{array}\right)\,.
\label{MatrixYY}
\end{equation}
This matrix can be obtained by considering the following linear combination
\begin{equation}
Y = \sum_{i}y_{i}\,T_{i}\,,
\end{equation}
where the six matrices $T_i$ appearing in the equation above are Lie algebra generators of $\mathfrak{so}(6)$ along the $\mathbb{CP}^3$ directions which can be written as follows
\begin{eqnarray}
T_1 = E_{13}-E_{31}-E_{24}+E_{42} \,, \qquad T_2 = E_{14}-E_{41}+E_{23}-E_{32}\,, \\
T_3 = E_{15}-E_{51}-E_{26}+E_{62} \,, \qquad T_4 = E_{16}-E_{61}+E_{25}-E_{52}\,, \\
T_5 = E_{35}-E_{53}-E_{46}+E_{64} \,, \qquad T_6 = E_{36}-E_{63}+E_{45}-E_{54}\,. 
\end{eqnarray}
Here $E_{ij}$ denote matrices  with a unit entry  in the $i^{th}$ row and $j^{th}$ column, all other entries being zero. The matrices $T_i$ are normalized as follows
\begin{equation}
\mbox{Tr}(T_iT_j) = -4\,\delta_{ij}\,.
\end{equation}
The nine generators of $\mathfrak{u}(3)$ inside $\mathfrak{so}(6)$ can be written as $[T_i,T_j]$. Of course only nine of these combinations are linearly independent.

Quite remarkably the matrix (\ref{MatrixYY}) obeys the following identities
\begin{equation}
Y^3 = -\rho^2 Y\,, \qquad \rho^2 = \sum^6_{i=1}\,y^2_i\,.
\label{IdentiCP3}
\end{equation}
Thus, a Lie algebra element $A$ parametrizing the coset space $AdS_4\times\mathbb{CP}^3$ can be represented in the following way
\begin{equation}
A = \left(\begin{array}{cc}
x_\mu\Gamma^\mu & 0 \\
0 & Y
\end{array}\right)\,, \qquad Y = \sum^6_{i=1}y_{i}\,T_{i}\,.
\end{equation}
Here $\Gamma^u$ are gamma matrices satisfying the Clifford algebra of $\mathfrak{so}(3,1)$.

In particular, an $SO(6)$ matrix parametrizing the coset space $SO(6)/U(3)$, and therefore $\mathbb{CP}^3$, can be obtained by exponentiating the generic element (\ref{MatrixYY}). The exponential of the matrix can be computed by using the identities (\ref{IdentiCP3}), and gives a coset representative of the following form
\begin{equation}
g = e^{Y} = I+\frac{\sin\rho}{\rho}\,Y+\frac{1-\cos\rho}{\rho^2}\,Y^2\,.
\end{equation}

\section{A basis for $\mathfrak{su}(4)$}

A basis for $\mathfrak{su}(4)$ can be constructed in terms of anti-Hermitian $4\times 4$ matrices. A direct generalization of the well known Gell-Mann matrices is given by
\begin{equation}
\begin{array}{ccc}
\lambda_1 = \left(\begin{array}{cccc}
0 & 1 & 0 & 0 \\
1 & 0 & 0 & 0 \\
0 & 0 & 0 & 0 \\
0 & 0 & 0 & 0
\end{array}
\right)\,,
&
\lambda_2 = \left(\begin{array}{cccc}
0 & -i & 0 & 0 \\
i & 0 & 0 & 0 \\
0 & 0 & 0 & 0 \\
0 & 0 & 0 & 0
\end{array}
\right)\,,
&
\lambda_3 = \left(\begin{array}{cccc}
1 & 0 & 0 & 0 \\
0 & -1 & 0 & 0 \\
0 & 0 & 0 & 0 \\
0 & 0 & 0 & 0
\end{array}
\right)\,,\\
& & \\
\lambda_4 = \left(\begin{array}{cccc}
0 & 0 & 1 & 0 \\
0 & 0 & 0 & 0 \\
1 & 0 & 0 & 0 \\
0 & 0 & 0 & 0
\end{array}
\right)\,,
&
\lambda_5 = \left(\begin{array}{cccc}
0 & 0 & -i & 0 \\
0 & 0 & 0 & 0 \\
i & 0 & 0 & 0 \\
0 & 0 & 0 & 0
\end{array}
\right)\,,
&
\lambda_6 = \left(\begin{array}{cccc}
0 & 0 & 0 & 0 \\
0 & 0 & 1 & 0 \\
0 & 1 & 0 & 0 \\
0 & 0 & 0 & 0
\end{array}
\right)\,,\\
& & \\
\lambda_7 = \left(\begin{array}{cccc}
0 & 0 & 0 & 0 \\
0 & 0 & -i & 0 \\
0 & i & 0 & 0 \\
0 & 0 & 0 & 0
\end{array}
\right)\,,
&
\lambda_8 = \frac{1}{\sqrt{3}}\left(\begin{array}{cccc}
1 & 0 & 0 & 0 \\
0 & 1 & 0 & 0 \\
0 & 0 & -2 & 0 \\
0 & 0 & 0 & 0
\end{array}
\right)\,,
&
\lambda_9 = \left(\begin{array}{cccc}
0 & 0 & 0 & 1 \\
0 & 0 & 0 & 0 \\
0 & 0 & 0 & 0 \\
1 & 0 & 0 & 0
\end{array}
\right)\,,\\
& & \\
\lambda_{10} = \left(\begin{array}{cccc}
0 & 0 & 0 & -i \\
0 & 0 & 0 & 0 \\
0 & 0 & 0 & 0 \\
i & 0 & 0 & 0
\end{array}
\right)\,,
&
\lambda_{11} = \left(\begin{array}{cccc}
0 & 0 & 0 & 0 \\
0 & 0 & 0 & 1 \\
0 & 0 & 0 & 0 \\
0 & 1 & 0 & 0
\end{array}
\right)\,,
&
\lambda_{12} = \left(\begin{array}{cccc}
0 & 0 & 0 & 0 \\
0 & 0 & 0 & -i \\
0 & 0 & 0 & 0 \\
0 & i & 0 & 0
\end{array}
\right)\,,\\
& & \\
\lambda_{13} = \left(\begin{array}{cccc}
0 & 0 & 0 & 0 \\
0 & 0 & 0 & 0 \\
0 & 0 & 0 & 1 \\
0 & 0 & 1 & 0
\end{array}
\right)\,,
&
\lambda_{14} = \left(\begin{array}{cccc}
0 & 0 & 0 & 0 \\
0 & 0 & 0 & 0 \\
0 & 0 & 0 & -i \\
0 & 0 & i & 0
\end{array}
\right)\,,
&
\lambda_{15} = \frac{1}{\sqrt{6}}\left(\begin{array}{cccc}
1 & 0 & 0 & 0 \\
0 & 1 & 0 & 0 \\
0 & 0 & 1 & 0 \\
0 & 0 & 0 & -3
\end{array}
\right)\,. \\
& & 
\end{array} 
\end{equation}
The first 8 matrices form a basis for $\mathfrak{su}(3)\subset \mathfrak{su}(4)$. Furthermore, these matrices are orthogonal and satisfy 
\begin{equation}
\mbox{Tr}\,(\lambda_m\,\lambda_n) = 2\,\delta_{mn}\,, \qquad m = 1,...\,, 15\, ,
\end{equation}
and commutation relations 
\begin{equation}
[\lambda_m,\lambda_n] = 2\,i\,f^{\,\,\,\,p}_{mn}\lambda_p\,.
\end{equation}
A list of non-vanishing structure constants can be found in \cite{Pfeifer}. In this representation the Cartan generators are given by $\lambda_3,\lambda_8$ and $\lambda_{15}$.

\acknowledgments

We would like to thank S. van Tongeren for valuable comments. The work of Ren\'e Negr\'on was supported by CAPES and the work of Victor O. Rivelles was supported by FAPESP grant 2014/18634-9.




\begin{thebibliography}{99}

\bibitem{Maldacena:1997re} 
  J.~M.~Maldacena,
  ``The Large N limit of superconformal field theories and supergravity,''
  \href{http://dx.doi.org/doi:10.1023/A:1026654312961}{Int.\ J.\ Theor.\ Phys.\  {\bf 38}, 1113 (1999)}
  \href{http://dx.doi.org/doi:10.4310/ATMP.1998.v2.n2.a1}{[Adv.\ Theor.\ Math.\ Phys.\  {\bf 2}, 231 (1998)]}
  \href{https://arxiv.org/abs/hep-th/9711200}{[hep-th/9711200]}.
  
\bibitem{BPR1} 
  I.~Bena, J.~Polchinski and R.~Roiban,
  ``Hidden symmetries of the AdS(5) x S**5 superstring,''
  \href{http://dx.doi.org/doi:10.1103/PhysRevD.69.046002}{Phys.\ Rev.\ D {\bf 69}, 046002 (2004)}
  \href{https://arxiv.org/abs/hep-th/0305116}{[hep-th/0305116]}.
  
\bibitem{Vicedo1} 
  B.~Vicedo,
  ``Hamiltonian dynamics and the hidden symmetries of the AdS(5) x S**5 superstring,''
  \href{http://dx.doi.org/doi:10.1007/JHEP01(2010)102}{JHEP {\bf 1001}, 102 (2010)}
  \href{https://arxiv.org/abs/0910.0221}{[arXiv:0910.0221 [hep-th]]}.
  
\bibitem{Magro1} 
  M.~Magro,
  ``The Classical Exchange Algebra of AdS(5) x S**5,''
  \href{http://dx.doi.org/doi:10.1088/1126-6708/2009/01/021}{JHEP {\bf 0901}, 021 (2009)}
  \href{https://arxiv.org/abs/0810.4136}{[arXiv:0810.4136 [hep-th]]}.

\bibitem{LuninMaldacena} 
  O.~Lunin and J.~M.~Maldacena,
  ``Deforming field theories with U(1) x U(1) global symmetry and their gravity duals,''
  \href{http://dx.doi.org/doi:10.1088/1126-6708/2005/05/033}{JHEP {\bf 0505}, 033 (2005)}
  \href{https://arxiv.org/abs/hep-th/0502086}{[hep-th/0502086]}.

\bibitem{Frolov2} 
  S.~A.~Frolov, R.~Roiban and A.~A.~Tseytlin,
  ``Gauge-string duality for superconformal deformations of N=4 super Yang-Mills theory,''
  \href{http://dx.doi.org/doi:10.1088/1126-6708/2005/07/045}{JHEP {\bf 0507}, 045 (2005)}
  \href{https://arxiv.org/abs/hep-th/0503192}{[hep-th/0503192]}.
  
\bibitem{Frolov3} 
  S.~Frolov,
  ``Lax pair for strings in Lunin-Maldacena background,''
  \href{http://dx.doi.org/doi:10.1088/1126-6708/2005/05/069}{JHEP {\bf 0505}, 069 (2005)}
  \href{https://arxiv.org/abs/hep-th/0503201}{[hep-th/0503201]}.

\bibitem{Imeroni:2008cr} 
  E.~Imeroni,
  ``On deformed gauge theories and their string/M-theory duals,''
  \href{http://dx.doi.org/doi:10.1088/1126-6708/2008/10/026}{JHEP {\bf 0810}, 026 (2008)}
  \href{https://arxiv.org/abs/0808.1271}{[arXiv:0808.1271 [hep-th]]}.

\bibitem{Vicedo6} 
  F.~Delduc, M.~Magro and B.~Vicedo,
  ``An integrable deformation of the $AdS_5 \times S^5$ superstring action,''
  \href{http://dx.doi.org/doi:10.1103/PhysRevLett.112.051601}{Phys.\ Rev.\ Lett.\  {\bf 112}, no. 5, 051601 (2014)}
  \href{https://arxiv.org/abs/1309.5850}{[arXiv:1309.5850 [hep-th]]}.

\bibitem{Klimcik} 
  C.~Klimcik,
  ``Yang-Baxter sigma models and dS/AdS T duality,''
  \href{http://dx.doi.org/doi:10.1088/1126-6708/2002/12/051}{JHEP {\bf 0212}, 051 (2002)}
  \href{https://arxiv.org/abs/hep-th/0210095}{[hep-th/0210095]}.

\bibitem{Klimcik2} 
  C.~Klimcik,
  ``On integrability of the Yang-Baxter sigma-model,''
  \href{http://dx.doi.org/doi:10.1063/1.3116242}{J.\ Math.\ Phys.\  {\bf 50}, 043508 (2009)}
  \href{https://arxiv.org/abs/0802.3518}{[arXiv:0802.3518 [hep-th]]}.

\bibitem{Arutyunov:2013ega} 
  G.~Arutyunov, R.~Borsato and S.~Frolov,
  ``S-matrix for strings on $\eta$-deformed AdS5 x S5,''
  \href{http://dx.doi.org/doi:10.1007/JHEP04(2014)002}{JHEP {\bf 1404}, 002 (2014)}
  \href{https://arxiv.org/abs/1312.3542}{[arXiv:1312.3542 [hep-th]]}.

\bibitem{Hoare:2014pna} 
  B.~Hoare, R.~Roiban and A.~A.~Tseytlin,
  ``On deformations of $AdS_n$ x $S^n$ supercosets,''
  \href{http://dx.doi.org/doi:10.1007/JHEP06(2014)002}{JHEP {\bf 1406}, 002 (2014)}
  \href{https://arxiv.org/abs/1403.5517}{[arXiv:1403.5517 [hep-th]]}.

\bibitem{Arutyunov:2015qva} 
  G.~Arutyunov, R.~Borsato and S.~Frolov,
  ``Puzzles of $\eta$-deformed AdS$_5 \times$ S$^5$,''
  \href{http://dx.doi.org/doi:10.1007/JHEP12(2015)049}{JHEP {\bf 1512}, 049 (2015)}
  \href{https://arxiv.org/abs/1507.04239}{[arXiv:1507.04239 [hep-th]]}.

\bibitem{Kawaguchi:2014qwa} 
  I.~Kawaguchi, T.~Matsumoto and K.~Yoshida,
  ``Jordanian deformations of the $AdS_5 \times S^5$ superstring,''
  \href{http://dx.doi.org/doi:10.1007/JHEP04(2014)153}{JHEP {\bf 1404}, 153 (2014)}
  \href{https://arxiv.org/abs/1401.4855}{[arXiv:1401.4855 [hep-th]]}.

\bibitem{Matsumoto:2014nra} 
  T.~Matsumoto and K.~Yoshida,
  ``Lunin-Maldacena backgrounds from the classical Yang-Baxter equation - Towards the gravity/CYBE correspondence,''
  \href{http://dx.doi.org/doi:10.1007/JHEP06(2014)135}{JHEP {\bf 1406}, 135 (2014)}
  \href{https://arxiv.org/abs/1404.1838}{[arXiv:1404.1838 [hep-th]]}.

\bibitem{Araujo:2017jkb} 
  T.~Araujo, I.~Bakhmatov, E.~\'O.~Colg\'ain, J.~Sakamoto, M.~M.~Sheikh-Jabbari and K.~Yoshida,
  ``Yang-Baxter $\sigma$-models, conformal twists, and noncommutative Yang-Mills theory,''
  \href{http://dx.doi.org/doi:10.1103/PhysRevD.95.105006}{Phys.\ Rev.\ D {\bf 95}, no. 10, 105006 (2017)}
  \href{https://arxiv.org/abs/1702.02861}{[arXiv:1702.02861 [hep-th]]}.
  
\bibitem{Bakhmatov:2017joy} 
  I.~Bakhmatov, \"O. Kelekci, E. \'O Colg\'ain and M.~M.~Sheikh-Jabbari,
  ``Classical Yang-Baxter Equation from Supergravity,''
  \href{http://dx.doi.org/doi:10.1103/PhysRevD.98.021901}{Phys.\ Rev.\ D {\bf 98}, no. 2, 021901 (2018)}
  \href{https://arxiv.org/abs/1710.06784}{[arXiv:1710.06784 [hep-th]]}.

\bibitem{Bakhmatov:2018apn} 
  I.~Bakhmatov, E. \'O. Colg\'ain, M.~M.~Sheikh-Jabbari and H.~Yavartanoo,
  ``Yang-Baxter Deformations Beyond Coset Spaces (a slick way to do TsT),''
  \href{http://dx.doi.org/doi:10.1007/JHEP06(2018)161}{JHEP {\bf 1806}, 161 (2018)}
  \href{https://arxiv.org/abs/1803.07498}{[arXiv:1803.07498 [hep-th]]}.
  
\bibitem{Matsumoto:2014gwa} 
  T.~Matsumoto and K.~Yoshida,
  ``Integrability of classical strings dual for noncommutative gauge theories,''
  \href{http://dx.doi.org/doi:10.1007/JHEP06(2014)163}{JHEP {\bf 1406}, 163 (2014)}
  \href{https://arxiv.org/abs/1404.3657}{[arXiv:1404.3657 [hep-th]]}.

\bibitem{Matsumoto} 
  T.~Matsumoto and K.~Yoshida,
  ``Integrable deformations of the AdS$_{5} \times S^5$ superstring and the classical Yang-Baxter equation $- Towards$ $the$ $gravity/CYBE$ $correspondence -$,''
  \href{http://dx.doi.org/doi:10.1088/1742-6596/563/1/012020}{J.\ Phys.\ Conf.\ Ser.\  {\bf 563}, no. 1, 012020 (2014)}
  \href{https://arxiv.org/abs/1410.0575}{[arXiv:1410.0575 [hep-th]]}.
  
\bibitem{Matsumoto:2014ubv} 
  T.~Matsumoto and K.~Yoshida,
  ``Yang-Baxter deformations and string dualities,''
  \href{http://dx.doi.org/doi:10.1007/JHEP03(2015)137}{JHEP {\bf 1503}, 137 (2015)}
  \href{https://arxiv.org/abs/1412.3658}{[arXiv:1412.3658 [hep-th]]}.

\bibitem{Matsumoto:2015uja} 
  T.~Matsumoto and K.~Yoshida,
  ``Schr\"odinger geometries arising from Yang-Baxter deformations,''
  JHEP {\bf 1504}, 180 (2015)
  \href{http://dx.doi.org/doi:10.1007/JHEP04(2015)180}{JHEP {\bf 1504}, 180 (2015)}
  \href{https://arxiv.org/abs/1502.00740}{[arXiv:1502.00740 [hep-th]]}.

\bibitem{vanTongeren:2015soa} 
  S.~J.~van Tongeren,
  ``On classical Yang-Baxter based deformations of the $AdS_{5}\times S^{5}$ superstring,''
  \href{http://dx.doi.org/doi:10.1007/JHEP06(2015)048}{JHEP {\bf 1506}, 048 (2015)}
  \href{https://arxiv.org/abs/1504.05516}{[arXiv:1504.05516 [hep-th]]}.

\bibitem{Matsumoto:2015ypa} 
  T.~Matsumoto, D.~Orlando, S.~Reffert, J.~i.~Sakamoto and K.~Yoshida,
  ``Yang-Baxter deformations of Minkowski spacetime,''
  \href{http://dx.doi.org/doi:10.1007/JHEP10(2015)185}{JHEP {\bf 1510}, 185 (2015)}
  \href{https://arxiv.org/abs/1505.04553}{[arXiv:1505.04553 [hep-th]]}.

\bibitem{Osten:2016dvf} 
  D.~Osten and S.~J.~van Tongeren,
  ``Abelian Yang-Baxter deformations and TsT transformations,''
  \href{http://dx.doi.org/doi:10.1016/j.nuclphysb.2016.12.007}{Nucl.\ Phys.\ B {\bf 915}, 184 (2017)}
  \href{https://arxiv.org/abs/1608.08504}{[arXiv:1608.08504 [hep-th]]}.

\bibitem{Hoare:2016wsk} 
  B.~Hoare and A.~A.~Tseytlin,
  ``Homogeneous Yang-Baxter deformations as non-abelian duals of the $AdS_5$ sigma-model,''
  \href{http://dx.doi.org/doi:10.1088/1751-8113/49/49/494001}{J.\ Phys.\ A {\bf 49}, no. 49, 494001 (2016)}
  \href{https://arxiv.org/abs/1609.02550}{[arXiv:1609.02550 [hep-th]]}.

\bibitem{Borsato:2016pas} 
  R.~Borsato and L.~Wulff,
  ``Integrable Deformations of $T$-Dual $\sigma$ Models,''
  \href{http://dx.doi.org/doi:10.1103/PhysRevLett.117.251602}{Phys.\ Rev.\ Lett.\  {\bf 117}, no. 25, 251602 (2016)}
  \href{https://arxiv.org/abs/1609.09834}{[arXiv:1609.09834 [hep-th]]}.
  
\bibitem{Borsato:2017qsx} 
  R.~Borsato and L.~Wulff,
  ``On non-abelian T-duality and deformations of supercoset string sigma-models,''
  \href{http://dx.doi.org/doi:10.1007/JHEP10(2017)024}{JHEP {\bf 1710}, 024 (2017)}
  \href{https://arxiv.org/abs/1706.10169}{[arXiv:1706.10169 [hep-th]]}.

\bibitem{Aharony:2008ug} 
  O.~Aharony, O.~Bergman, D.~L.~Jafferis and J.~Maldacena,
  ``N=6 superconformal Chern-Simons-matter theories, M2-branes and their gravity duals,''
  \href{http://dx.doi.org/doi:10.1088/1126-6708/2008/10/091}{JHEP {\bf 0810}, 091 (2008)}
  \href{https://arxiv.org/abs/0806.1218}{[arXiv:0806.1218 [hep-th]]}.

\bibitem{Arutyunov:2008if} 
  G.~Arutyunov and S.~Frolov,
  ``Superstrings on $AdS_4 \times CP^3$ as a Coset Sigma-model,''
  \href{http://dx.doi.org/doi:10.1088/1126-6708/2008/09/129}{JHEP {\bf 0809}, 129 (2008)}
  \href{https://arxiv.org/abs/0806.4940}{[arXiv:0806.4940 [hep-th]]}.

\bibitem{Klose:2010ki} 
  T.~Klose,
  ``Review of AdS/CFT Integrability, Chapter IV.3: N=6 Chern-Simons and Strings on AdS4xCP3,''
  \href{http://dx.doi.org/doi:10.1007/s11005-011-0520-y}{Lett.\ Math.\ Phys.\  {\bf 99}, 401 (2012)}
  \href{https://arxiv.org/abs/1012.3999}{[arXiv:1012.3999 [hep-th]]}.

\bibitem{Crichigno:2014ipa} 
  P.~M.~Crichigno, T.~Matsumoto and K.~Yoshida,
  ``Deformations of $T^{1,1}$ as Yang-Baxter sigma models,''
  \href{http://dx.doi.org/doi:10.1007/JHEP12(2014)085}{JHEP {\bf 1412}, 085 (2014)}
  \href{https://arxiv.org/abs/1406.2249}{[arXiv:1406.2249 [hep-th]]}.

\bibitem{Zarembo:2010sg} 
  K.~Zarembo,
  ``Strings on Semisymmetric Superspaces,''
  \href{http://dx.doi.org/doi:10.1007/JHEP05(2010)002}{JHEP {\bf 1005}, 002 (2010)}
  \href{https://arxiv.org/abs/1003.0465}{[arXiv:1003.0465 [hep-th]]}.
  
\bibitem{Stefanski:2008ik} 
  B.~Stefanski, jr,
  ``Green-Schwarz action for Type IIA strings on AdS(4) x CP**3,''
  \href{http://dx.doi.org/doi:10.1016/j.nuclphysb.2008.09.015}{Nucl.\ Phys.\ B {\bf 808}, 80 (2009)}
  \href{https://arxiv.org/abs/0806.4948}{[arXiv:0806.4948 [hep-th]]}.
  
\bibitem{Sorokin:2010wn} 
  D.~Sorokin and L.~Wulff,
  ``Evidence for the classical integrability of the complete $AdS_4 x CP^3$ superstring,''
  \href{http://dx.doi.org/doi:10.1007/JHEP11(2010)143}{JHEP {\bf 1011}, 143 (2010)}
  \href{https://arxiv.org/abs/1009.3498}{[arXiv:1009.3498 [hep-th]]}.
  
\bibitem{Gomis:2008jt} 
  J.~Gomis, D.~Sorokin and L.~Wulff,
  ``The Complete AdS(4) x CP**3 superspace for the type IIA superstring and D-branes,''
  \href{http://dx.doi.org/doi:10.1088/1126-6708/2009/03/015}{JHEP {\bf 0903}, 015 (2009)}
  \href{https://arxiv.org/abs/0811.1566}{[arXiv:0811.1566 [hep-th]]}.

\bibitem{Pfeifer}
  Walter Pfeifer,
  ``The Lie algebras su(N), An Introduction,''
  \href{https://www.springer.com/us/book/9783764324186}{Birkh\"ausser Bassel, 2003}.

\bibitem{Vicedo5} 
  F.~Delduc, M.~Magro and B.~Vicedo,
  ``On classical $q$-deformations of integrable sigma-models,''
  \href{http://dx.doi.org/doi:10.1007/JHEP11(2013)192}{JHEP {\bf 1311}, 192 (2013)}
  \href{https://arxiv.org/abs/1308.3581}{[arXiv:1308.3581 [hep-th]]}.
  
\bibitem{Matsumoto:2015jja} 
  T.~Matsumoto and K.~Yoshida,
  ``Yang-Baxter sigma models based on the CYBE,''
  \href{http://dx.doi.org/doi:10.1016/j.nuclphysb.2015.02.009}{Nucl.\ Phys.\ B {\bf 893}, 287 (2015)}
  \href{https://arxiv.org/abs/1501.03665}{[arXiv:1501.03665 [hep-th]]}.

\bibitem{Borsato:2016ose} 
  R.~Borsato and L.~Wulff,
  ``Target space supergeometry of $\eta$ and $\lambda$-deformed strings,''
  \href{http://dx.doi.org/doi:10.1007/JHEP10(2016)045}{JHEP {\bf 1610}, 045 (2016)}
  \href{https://arxiv.org/abs/1608.03570}{[arXiv:1608.03570 [hep-th]]}.

\bibitem{vanTongeren:2015uha} 
  S.~J.~van Tongeren,
  ``Yang-Baxter deformations, AdS/CFT, and twist-noncommutative gauge theory,''
  \href{http://dx.doi.org/doi:10.1016/j.nuclphysb.2016.01.012}{Nucl.\ Phys.\ B {\bf 904}, 148 (2016)}
  \href{https://arxiv.org/abs/1506.01023}{[arXiv:1506.01023 [hep-th]]}.

\bibitem{vanTongeren:2016eeb} 
  S.~J.~van Tongeren,
  ``Almost abelian twists and AdS/CFT,''
  \href{http://dx.doi.org/doi:10.1016/j.physletb.2016.12.002}{Phys.\ Lett.\ B {\bf 765}, 344 (2017)}
  \href{https://arxiv.org/abs/1610.05677}{[arXiv:1610.05677 [hep-th]]}.

\bibitem{Kyono:2016jqy} 
  H.~Kyono and K.~Yoshida,
  ``Supercoset construction of Yang-Baxter deformed AdS$_5\times$S$^5$ backgrounds,''
  \href{http://dx.doi.org/doi:10.1093/ptep/ptw111}{PTEP {\bf 2016}, no. 8, 083B03 (2016)}
  \href{https://arxiv.org/abs/1605.02519}{[arXiv:1605.02519 [hep-th]]}.



  



  




\end{thebibliography}
\end{document}